\documentclass[12pt,a4paper]{article}
\pdfoutput=1

\usepackage{amssymb,amsmath,authblk,hyperref}
\usepackage{fullpage}

\DeclareMathOperator{\diag}{diag}
\DeclareMathOperator{\tr}{tr}

\title{Physical remnant of electroweak theta angles}

\author[1]{James~Brister\footnote{\texttt{jbrister@scu.edu.cn}}}
\author[1,2]{Bingwei~Long\footnote{\texttt{bingwei@scu.edu.cn}}}
\author[1]{Longjie~Ran\footnote{\texttt{2019222020004@stu.scu.edu.cn}}}
\author[1]{Muhammad~Shahzad\footnote{\texttt{2023521221001@stu.scu.edu.cn}}}
\author[1]{Zheng~Sun\footnote{\texttt{sun\_ctp@scu.edu.cn,}}}
\author[1]{Yingpei~Zou\footnote{\texttt{yingpei@stu.scu.edu.cn}}}

\affil[1]{College of Physics, Sichuan University, \protect \\
          Chengdu 610065, Sichuan Province, P.\ R.\ China}
\affil[2]{Southern Center for Nuclear-Science Theory (SCNT), \protect \\
          Institute of Modern Physics, Chinese Academy of Sciences, \protect \\
          Huizhou 516000, Guangdong Province, P.\ R.\ China}

\date{}

\begin{document}

\maketitle

\begin{abstract}
In addition to the well-known quantum chromodynamical theta angle, we show that the Standard Model has another theta angle which is invariant under arbitrary chiral rotations of quarks and leptons.  The new theta angle can be identified with the quantum electrodynamical theta angle, which should be viewed as an independent parameter of the Standard Model, and may be observable in spacetime with non-simply connected features, either beyond the visible universe or in an effective background from a laboratory setup.
\end{abstract}

\section{Introduction}

The charge-parity (CP) violating theta terms can naturally be included in the Standard Model (SM), since the CP symmetry has already been violated in the Yukawa coupling sector~\cite{Branco:1999fs, Bigi:2000yz}.  The effective quantum chromodynamical (QCD) theta angle is constrained to an unnatural limit $\lvert \bar \theta_\text{QCD} \rvert < 10^{- 10}$ by the current measurements of the neutron electric dipole moment (EDM)~\cite{ParticleDataGroup:2024cfk}.  One solution to this ``strong CP problem'' is to introduce the axion which may be identified as a dark matter candidate~\cite{Dine:2000cj, Peccei:2006as, Kim:2008hd, Hook:2018dlk, DiLuzio:2020wdo}.  The weak $SU(2)$ theta angle can be removed by a chiral rotation of quarks and leptons which does not modify their masses, unless beyond the SM operators are considered to prevent the removal~\cite{Anselm:1992yz, Anselm:1993uj, FileviezPerez:2014xju, Shifman:2017lkj}.  A $U(1)$ theta angle has no physical effect because of the absence of $U(1)$ instantons in Minkowski spacetime.  Therefore theta terms in the electroweak sector are usually considered unphysical.

On the other hand, the $U(1)$ theta angle may become physical if the gauge theory is formulated on a compact and non-simply connected spacetime manifold~\cite{Witten:1995gf, Verlinde:1995mz, Olive:2000yy}.  Such a construction can be realized by considering nontrivial topological features beyond the visible universe, or in an effective background from a laboratory setup~\cite{Hsu:2010jm, Hsu:2011sx, Cao:2013na, Cao:2017ocv, Zhitnitsky:2023jvp}.  An effective $U(1)$ theta angle $\bar \theta_\text{QED}$ can be included in quantum electrodynamics (QED), since it is invariant under a vector rotation of quarks and leptons.  One may ask: is $\bar \theta_\text{QED}$ a remnant of the SM $SU(2)$ and $U(1)$ theta angles after electroweak symmetry breaking (EWSB), or must it be introduced as an explicit symmetry breaking parameter?  In what follows, we will show that besides the well-known $\bar \theta_\text{QCD}$, there is another combination of the SM theta angles which are invariant under any chiral rotation of quarks and leptons.  This new invariant combination can be identified with $\bar \theta_\text{QED}$, which should be viewed as an independent parameter of the SM, and may lead to physical observables measurable by experiments.

\section{Chiral rotations and shifts of theta angles}

A chiral $U(1)$ rotates left- and right-handed fermions with independent phases, i.e.:
\begin{equation}
\psi_L \to e^{i \alpha_{\psi L}} \psi_L, \quad
\psi_R \to e^{- i \alpha_{\psi R}} \psi_R,
\end{equation}
where $\psi_L$ and $\psi_R$ could be either two-component Weyl fermions or chiral fermions in four-component notation~\cite{Dreiner:2008tw, Dreiner:2023yus}.  One can consistently write
\begin{equation}
\psi \to e^{- i \gamma^5 \alpha_\psi} \psi, \quad
\gamma^5 = \diag(- \mathbf{1}^{2 \times 2},
                 \mathbf{1}^{2 \times 2}) \label{eq:2-01}
\end{equation}
for $\psi$ being either left- or right-handed fermions.  If $\psi_L$ and $\psi_R$ have the same set of quantum numbers, they can appear in the mass term $- m \bar \psi_R \psi_L$.  A general chiral rotation modifies the mass as
\begin{equation}
m \to e^{i (\alpha_{\psi L} + \alpha_{\psi R})} m, \quad
\phi = \arg m \to \phi + \alpha_{\psi L} + \alpha_{\psi R}. \label{eq:2-02}
\end{equation}
Specifically, $\alpha_{\psi L} = - \alpha_{\psi R}$ corresponds to a vector rotation which keeps the mass invariant, and $\alpha_{\psi L} = \alpha_{\psi R}$ corresponds to an axial rotation which modifies the mass by a phase $2 \alpha_{\psi L}$.

When a massless chiral fermion $\psi$ couples to a gauge field $A_\mu$, the classically conserved chiral current
\begin{equation}
j^{\mu 5} = \bar \psi \gamma^\mu \gamma^5 \psi
\end{equation}
becomes quantumly non-conserved due to the chiral anomaly.  The divergence of the chiral current can be calculated from the fermion functional measure change in the path integral, or from triangle Feynman diagrams attached to $j^{\mu 5}$.  The result is
\begin{equation}
\partial_\mu j^{\mu 5} = - \frac{g^2}{32 \pi^2}
                           \epsilon^{\mu \nu \kappa \lambda}
                           \tr_\psi (F_{\mu \nu} F_{\kappa \lambda})
                       = - \partial_\mu K^\mu,
\end{equation}
where $g$ is the gauge coupling constant, $K^\mu$ is the Chern-Simons current
\begin{equation}
K^\mu = \frac{g^2}{16 \pi^2} \epsilon^{\mu \nu \kappa \lambda}
        \tr_\psi (A_\nu F_{\kappa \lambda}
                  - \frac{i g}{3} [A_\nu, A_\kappa] A_\lambda),
\end{equation}
and the trace $\tr_\psi$ is applied to the gauge group representation matrix on the fermion $\psi$.  The conserved but gauge-dependent current $\bar j^{\mu 5} = j^{\mu 5} + K^\mu$ can be identified as the Noether current of the chiral symmetry, which indicates that the Lagrangian changes by a four-divergence under the chiral rotation~\eqref{eq:2-01}, i.e.,
\begin{equation}
\alpha_\psi \Delta \mathcal{L}
= - \alpha_\psi \partial_\mu K^\mu
= - \frac{\alpha_\psi g^2}{32 \pi^2}
    \epsilon^{\mu \nu \kappa \lambda}
    \tr_\psi (F_{\mu \nu} F_{\kappa \lambda}). \label{eq:2-03}
\end{equation}
This result can be compared with the theta term
\begin{equation}
\mathcal{L}_\theta = \frac{\theta g^2}{32 \pi^2}
                     \epsilon^{\mu \nu \kappa \lambda}
                     \tr_F (F_{\mu \nu} F_{\kappa \lambda}) \label{eq:2-04}
\end{equation}
where the trace $\tr_F$ is applied to the gauge fields viewed either as vectors in the adjoint representation space, or as the representation matrices on the fundamental representation space.  We see the shift of the theta angle
\begin{equation}
\theta \to \theta - \frac{I_\psi}{I_F} \alpha_\psi \label{eq:2-05}
\end{equation}
under the chiral rotation~\eqref{eq:2-01}, where the second-order indices $I_\psi$ and $I_F$ are determined by
\begin{equation}
\tr (R(t_a) R(t_b)) = I_R g_{a b}
\end{equation}
for an irreducible representation (irrep) $R$.  For a simple Lie group, $I_R$ is identified as the quadratic Dynkin index of the irrep $R$~\cite{Slansky:1981yr, Ramond:2010zz}.  The Killing metric $g_{a b}$ is independent of representations, but depends on the choice of generators $t_a$.  The typical convention in physics sets $g_{a b} = \delta_{a b}$~\cite{Peskin:1995ev, Schwartz:2014sze}, so the traces in~\eqref{eq:2-03} and~\eqref{eq:2-04} become
\begin{equation}
\tr_R (F_{\mu \nu} F_{\kappa \lambda})
= \tr (R(F_{\mu \nu}^a t_a) R(F_{\kappa \lambda}^b t_b))
= I_R F_{\mu \nu}^a F_{\kappa \lambda}^a,
\end{equation}
and we have $I_0 = 0$, $I_F = \frac{1}{2}$ and $I_{Ad} = N$ for trivial, fundamental and adjoint representations of $SU(N)$.  The same convention $g_{a b} = \delta_{a b}$ for $U(1)$ leads to $I_q = q^2$ for a charge-$q$ representation, and the fundamental representation corresponds to $q = 1$.

\section{Chiral rotations in the Standard Model}

The effects \eqref{eq:2-02} and \eqref{eq:2-05} enable us to remove unphysical theta angles from the SM by applying a chiral rotation.  There are three theta terms for the SM gauge group $SU(3) \times SU(2) \times U(1)$:
\begin{equation}
\mathcal{L}_\theta = \sum_{n=1}^3
                     \frac{\theta_n g_n^2}{32 \pi^2}
                     \epsilon^{\mu \nu \kappa \lambda}
                     \tr_F (F^{(n)}_{\mu \nu}
                           F^{(n)}_{\kappa \lambda})
                   = \sum_{n=1}^3
                     \frac{\theta_n g_n^2 I^{(n)}_F}{32 \pi^2}
                     \epsilon^{\mu \nu \kappa \lambda}
                     F^{(n) a}_{\mu \nu}
                     F^{(n) a}_{\kappa \lambda}, \label{eq:3-01}
\end{equation}
where $n = 1, 2, 3$ labels each indecomposable subgroup $U(1)$, $SU(2)$ or $SU(3)$.  The SM fermions correspond to gauge group irreps $(R_3, R_2)_Y$, where $R_3$ and $R_2$ are $SU(3)$ and $SU(2)$ irreps labeled by their dimensions, and $Y$ is the $U(1)$ hypercharge, i.e.,
\begin{gather}
q_L^i = (u_L^i, d_L^i)^T \sim (3, 2)_{1 / 6}, \quad
u_R^i \sim (3, 1)_{2 / 3}, \quad
d_R^i \sim (3, 1)_{- 1 / 3}, \label{eq:3-02}\\ 
l_L^i = (\nu_L^i, e_L^i)^T \sim (1, 2)_{- 1 / 2}, \quad
e_R^i \sim (1, 1)_{- 1}, \label{eq:3-03}
\end{gather}
where $i = 1, 2, 3$ identifies three generations of quarks and leptons.  They appear in fermion mass terms after electroweak symmetry breaking (EWSB):
\begin{equation}
\mathcal{L}_{m f} = - (M_u^{i j} \bar u_R^j u_L^i
                       + M_d^{i j} \bar d_R^j d_L^i
                       + M_e^{i j} \bar e_R^j e_L^i
                       + \text{h.c.}).
\end{equation}
Neutrinos can have either Dirac mass terms
\begin{equation}
\mathcal{L}_{m \nu} = - (M_\nu^{i j} \bar \nu_R^j \nu_L^i
                         + \text{h.c.}) \label{eq:3-04}
\end{equation}
after EWSB by introducing sterile right-handed neutrinos $\nu_R^i \sim (1, 1)_0$, or effective Majorana mass terms
\begin{equation}
\mathcal{L}_{m \nu} = - \frac{1}{2}
                        (M_\nu^{i j} (E^{- 1} \nu_L^i)^T \nu_L^i
                         + \text{h.c.}) \label{eq:3-05}
\end{equation}
from seesaw mechanism or loop contributions, where $E^{- 1} = \diag(\epsilon^{(2)}, - \epsilon^{(2)})$ is the metric for raising and lowering spinor indices, with $\epsilon^{(2)}$ being the rank-two Levi-Civita symbol.  Chiral rotations of fermions contribute to the shifts of theta angles as well as phases of fermion masses according to \eqref{eq:2-02} and \eqref{eq:2-05}.

There are $12$ independent phases in total for chiral rotations in the quark sector.  Reducing complex phases in the Cabibbo-Kobayashi-Maskawa (CKM) matrix from $6$ to $1$ without shifting theta angles fixes $5$ of $6$ rotations of $q_L^i$.  The remaining $1$ rotation by a common phase $\alpha_{q L}$ for $q_L^i$ does not change the CKM matrix but shifts theta angles.  Then setting all quark masses to real values fixes all $6$ rotations of $u_R^i$ and $d_R^i$.  We can separate out $3$ common phases $\alpha_{q L}$, $\alpha_{u R}$ and $\alpha_{d R}$, which modify the overall phases of $M_u^{i j}$ and $M_d^{i j}$ without changing the CKM matrix:
\begin{align}
\phi_u &= \arg \det M_u^{i j}
        \to \phi_u + 3 (\alpha_{q L} + \alpha_{u R}), \label{eq:3-06}\\
\phi_d &= \arg \det M_d^{i j}
        \to \phi_d + 3 (\alpha_{q L} + \alpha_{d R}).
\end{align}
Once $\phi_u$ and $\phi_d$ are set to zero, the remaining $4$ rotations between different generations of $u_R^i$ or $d_R^i$ set the individual masses to reals without shifting theta angles.  The same argument goes through in the lepton sector if neutrinos have only Dirac mass terms as~\eqref{eq:3-04}, and we can separate out $\alpha_{l L}$, $\alpha_{e R}$ and $\alpha_{\nu R}$ to modify the overall phases of $M_e^{i j}$ and $M_\nu^{i j}$ without changing the Pontecorvo-Maki-Nakagawa-Sakata (PMNS) matrix:
\begin{align}
\phi_e &= \arg \det M_e^{i j}
        \to \phi_e + 3 (\alpha_{l L} + \alpha_{e R}), \label{eq:3-07}\\
\phi_\nu &= \arg \det M_\nu^{i j}
          \to \phi_\nu + 3 (\alpha_{l L} + \alpha_{\nu R}). \label{eq:3-08}
\end{align}
In the case that the SM neutrinos have effective Majorana mass terms as~\eqref{eq:3-05}, we have only $\alpha_{l L}$ and $\alpha_{e R}$ to modify the overall phases of $M_e^{i j}$ and $M_\nu^{i j}$, and \eqref{eq:3-08} should be replaced by
\begin{equation}
\phi_\nu \to \phi_\nu + 6 \alpha_{l L}.
\end{equation}
In summary, we consider the chiral rotation
\begin{gather}
q_L^i \to e^{i \alpha_{q L}} q_L^i, \quad
u_R^i \to e^{- i \alpha_{u R}} u_R^i, \quad
d_R^i \to e^{- i \alpha_{d R}} d_R^i, \label{eq:3-09}\\
l_L^i \to e^{i \alpha_{l L}} l_L^i, \quad
e_R^i \to e^{- i \alpha_{e R}} e_R^i, \quad
\nu_R^i \to e^{- i \alpha_{\nu R}} \nu_R^i \label{eq:3-10}
\end{gather}
specified by the phases $\alpha_{q L}$, $\alpha_{u R}$, $\alpha_{d R}$, $\alpha_{l L}$, $\alpha_{e R}$ and $\alpha_{\nu R}$, which commutes with the SM gauge symmetry, and modifies only the overall phases of SM fermions listed in~\eqref{eq:3-02} and~\eqref{eq:3-03} as well as the right-handed neutrinos $\nu_R^i$ in the Dirac neutrino mass terms~\eqref{eq:3-04}.

\section{Invariant combinations of theta angles}

The shifts of theta angles under a chiral rotation get contributions from all fermions.  Each fermion $\psi$ in an irrep of the $n$-th gauge subgroup shifts $\theta_n$ according to~\eqref{eq:2-05}.  This shift is multiplied by the total number of fermions in the same irrep, including the number of generations $N^{(f)}_\psi = 3$, and the dimensions $d^{(n' \ne n)}_\psi$ of the $n'$-th gauge subgroup irreps on $\psi$, i.e.,
\begin{equation}
\theta_n \to \theta_n - \frac{1}{I^{(n)}_F}
                        \sum_\psi I^{(n)}_\psi \alpha_\psi
                                  N^{(f)}_\psi
                                  \prod_{n' \ne n} d^{(n')}_\psi.
\end{equation}
Summing up all contributions from the SM fermions listed in~\eqref{eq:3-02} and~\eqref{eq:3-03}, we obtain the shifts of theta angles under the chiral rotation~\eqref{eq:3-09} and~\eqref{eq:3-10}:
\begin{align}
\theta_3 &\to \theta_3
              - (6 \alpha_{q L} 
                 + 3 \alpha_{u R} + 3 \alpha_{d R}), \label{eq:4-01}\\
\theta_2 &\to \theta_2
              - (9 \alpha_{q L} + 3 \alpha_{l L}),\\
\theta_1 &\to \theta_1
              - (\frac{1}{2} \alpha_{q L} + 4 \alpha_{u R} + \alpha_{d R}
                 + \frac{3}{2} \alpha_{l L} + 3 \alpha_{e R}). \label{eq:4-02}
\end{align}

From the transformation properties~\eqref{eq:3-06}--\eqref{eq:3-07} and~\eqref{eq:4-01}--\eqref{eq:4-02}, we find two combinations
\begin{align}
\bar \theta_3 &= \theta_3 + \phi_u + \phi_d, \label{eq:4-03}\\
\bar \theta_{2 1} &= \frac{1}{2} \theta_2 + \theta_1
                     + \frac{4}{3} \phi_u + \frac{1}{3} \phi_d + \phi_e \label{eq:4-04}
\end{align}
which are invariant under arbitrary chiral rotations of fermions.  One can apply a chiral rotation with
\begin{align}
\alpha_{u R} &= - \alpha_{q L} - \frac{1}{3} \phi_u, \label{eq:4-05}\\
\alpha_{d R} &= - \alpha_{q L} - \frac{1}{3} \phi_d, \label{eq:4-06}\\
\alpha_{e R} &= - \alpha_{l L} - \frac{1}{3} \phi_e \label{eq:4-07}
\end{align}
to shift $\phi_u$, $\phi_d$ and $\phi_e$ to zero, so that the nonzero values of $\bar \theta_3$ and $\bar \theta_{2 1}$ are allocated to $\theta_3$ and $\frac{1}{2} \theta_2 + \theta_1$.  The value of $\alpha_{l L}$ can be chosen to shift $\phi_\nu$ to zero: we have
\begin{equation}
\alpha_{\nu R} = - \alpha_{l L} - \frac{1}{3} \phi_\nu \label{eq:4-08}
\end{equation}
for $M_\nu^{i j}$ being Dirac, and
\begin{equation}
\alpha_{l L} = - \frac{1}{6} \phi_\nu \label{eq:4-09}
\end{equation}
for $M_\nu^{i j}$ being Majorana.  The remaining degree of freedom in \eqref{eq:4-05} and \eqref{eq:4-06} parameterizes the baryon number symmetry $U(1)_B$ with
\begin{equation}
\alpha(B) = 3 \Delta \alpha_{q L}
          = - 3 \Delta \alpha_{u R}
          = - 3 \Delta \alpha_{d R}.
\end{equation}
This degree of freedom can be used to remove $\theta_2$ by choosing
\begin{equation}
\alpha_{q L} = \frac{1}{3} \alpha(B)
             = \frac{1}{9} \theta_2 - \frac{1}{3} \alpha_{l L}, \label{eq:4-10}
\end{equation}
so that the nonzero value of $\bar \theta_{2 1}$ can be allocated to $\theta_1$.  Similarly, the remaining degree of freedom in \eqref{eq:4-07} and \eqref{eq:4-08} parameterizes the lepton number symmetry $U(1)_L$ with
\begin{equation}
\alpha(L) = \Delta \alpha_{l L}
          = - \Delta \alpha_{e R}
          = - \Delta \alpha_{\nu R}
\end{equation}
if the neutrino masses are Dirac.  While $\theta_2$ and $\theta_1$ are not invariant under $U(1)_B$ and $U(1)_L$, all of $\theta_3$, $\theta_2$ and $\theta_1$ are invariant under $U(1)_{B - L}$ which is parameterized by
\begin{equation}
\alpha(B - L) = 3 \Delta \alpha_{q L}
              = - 3 \Delta \alpha_{u R}
              = - 3 \Delta \alpha_{d R}
              = - \Delta \alpha_{l L}
              = \Delta \alpha_{e R}
              = \Delta \alpha_{\nu R}.
\end{equation}
If neutrinos have Majorana masses which violate $U(1)_L$, then $\alpha_{l L}$ is fixed by \eqref{eq:4-09} and there is no remaining degree of freedom in the lepton sector.  Because of the absence of $\phi_\nu$ in~\eqref{eq:4-03} and~\eqref{eq:4-04}, the existence and invariance of $\bar \theta_3$ and $\bar \theta_{2 1}$ is independent of whether neutrinos have Dirac or Majorana masses.

\section{Physical theta angles in the Standard Model}

To see the physical meaning of $\bar \theta_3$ and $\bar \theta_{2 1}$, we convert the electroweak gauge fields $W^A_\mu$ and $B_\mu$ to fields $W^\pm_\mu$, $Z_\mu$ and $A_\mu$ after EWSB:
\begin{gather}
W^1_\mu = \frac{1}{\sqrt{2}} (W^+_\mu + W^-_\mu),\quad
W^2_\mu = \frac{i}{\sqrt{2}} (W^+_\mu - W^-_\mu),\\
W^3_\mu = \frac{1}{\sqrt{g_2^2 + g_1^2}} (g_2 Z_\mu + g_1 A_\mu), \quad
B_\mu = \frac{1}{\sqrt{g_2^2 + g_1^2}} (- g_1 Z_\mu + g_2 A_\mu).
\end{gather}
Using these relations, the electroweak field strengths
\begin{equation}
F^{(2) A}_{\mu \nu} = \partial_\mu W^A_\nu - \partial_\nu W^A_\mu
                      - g_2 \epsilon^{A B C} W^B_\mu W^C_\nu, \quad
F^{(1)}_{\mu \nu} = \partial_\mu B_\nu - \partial_\nu B_\mu
\end{equation}
are expressed in terms of $W^\pm_\mu$, $Z_\mu$ and $A_\mu$.  Then the SM theta terms~\eqref{eq:3-01} become
\begin{equation}
\begin{split}
\mathcal{L}_\theta &= \frac{\theta_3 g_3^2}{64 \pi^2}
                      \epsilon^{\mu \nu \kappa \lambda}
                      G^a_{\mu \nu} G^a_{\kappa \lambda}
                      + (\frac{1}{2} \theta_2 + \theta_1)
                        \frac{e^2}{32 \pi^2}
                        \epsilon^{\mu \nu \kappa \lambda}
                        A_{\mu \nu} A_{\kappa \lambda}\\
                   &\hphantom{\mbox{} = \mbox{}}
                      + (\frac{1}{2} \theta_2  \cot^4 \theta_W + \theta_1)
                        \frac{e^2}{32 \pi^2} \tan^2 \theta_W
                        \epsilon^{\mu \nu \kappa \lambda}
                        Z_{\mu \nu} Z_{\kappa \lambda}\\
                   &\hphantom{\mbox{} = \mbox{}}
                      + (\theta_2 \cot^2 \theta_W - 2 \theta_1)
                        \frac{e^2}{32 \pi^2} \tan \theta_W
                        \epsilon^{\mu \nu \kappa \lambda}
                        A_{\mu \nu} Z_{\kappa \lambda}\\
                   &\hphantom{\mbox{} = \mbox{}}
                      + \frac{\theta_2 e^2 \csc^2 \theta_W}{32 \pi^2}
                        \epsilon^{\mu \nu \kappa \lambda}\\
                   &\hphantom{\mbox{} = \mbox{} + \mbox{}}
                        \times ((\partial_\mu W^+_\nu - \partial_\nu W^+_\mu)
                                (\partial_\kappa W^-_\lambda
                                 - \partial_\lambda W^-_\kappa)\\
                   &\hphantom{\mbox{} = \mbox{} + \mbox{} \times (}
                                + 4 i e
                                  (\partial_\mu A_\nu W^+_\kappa W^-_\lambda
                                   + A_\nu \partial_\mu W^+_\kappa W^-_\lambda
                                   + A_\nu W^+_\kappa \partial_\mu W^-_\lambda\\
                   &\hphantom{\mbox{} = \mbox{} + \mbox{} \times (
                                        \mbox{} + 4 i e (}
                                   + \cot \theta_W
                                     (\partial_\mu Z_\nu W^+_\kappa W^-_\lambda
                                      + Z_\nu \partial_\mu W^+_\kappa W^-_\lambda
                                      + Z_\nu W^+_\kappa \partial_\mu W^-_\lambda)))\\
                   &= \frac{1}{8 \pi^2} \epsilon^{\mu \nu \kappa \lambda}
                      \partial_\mu (\frac{1}{2} \theta_3 g_3^2
                                    (G^a_\nu \partial_\kappa G^a_\lambda
                                     - \frac{2}{3} g_3 f^{a b c}
                                       G^a_\nu G^b_\kappa G^c_\lambda)
                                    + (\frac{1}{2} \theta_2 + \theta_1)
                                      e^2
                                      A_\nu \partial_\kappa A_\lambda\\
                   &\hphantom{\mbox{} = \frac{1}{8 \pi^2}
                                        \epsilon^{\mu \nu \kappa \lambda}
                                        \partial_\mu (}
                                    + (\frac{1}{2} \theta_2 \cot^4 \theta_W
                                       + \theta_1)
                                      e^2 \tan^2 \theta_W
                                      Z_\nu \partial_\kappa Z_\lambda\\
                   &\hphantom{\mbox{} = \frac{1}{8 \pi^2}
                                        \epsilon^{\mu \nu \kappa \lambda}
                                        \partial_\mu (}
                                    + (\theta_2 \cot^2 \theta_W - 2 \theta_1)
                                      e^2 \tan \theta_W
                                      A_\nu \partial_\kappa Z_\lambda\\
                   &\hphantom{\mbox{} = \frac{1}{8 \pi^2}
                                        \epsilon^{\mu \nu \kappa \lambda}
                                        \partial_\mu (}
                                    + \theta_2 e^2 \csc^2 \theta_W
                                      (W^+_\nu \partial_\kappa W^-_\lambda
                                       + i e 
                                         (A_\nu W^+_\kappa W^-_\lambda
                                          + \cot \theta_W Z_\nu W^+_\kappa W^-_\lambda))),
\end{split} \label{eq:5-01}
\end{equation}
where $G^a_{\mu \nu} = F^{(3) a}_{\mu \nu} = \partial_\mu G^a_\nu - \partial_\nu G^a_\mu - g_3 f^{a b c} G^b_\mu G^c_\nu$ is the field strength of the gluon field $G^a_\mu$, $A_{\mu \nu} = \partial_\mu A_\nu - \partial_\nu A_\mu$ is the field strength of the electromagnetic field $A_\mu$, $Z_{\mu \nu} = \partial_\mu Z_\nu - \partial_\nu Z_\mu$ is the field strength of $Z_\mu$, $e = g_2 g_1 / \sqrt{g_2^2 + g_1^2}$ is the elementary charge or the electromagnetic coupling constant, and $\theta_W = \arctan (g_1 / g_2)$ is the weak mixing angle.  It can be seen in \eqref{eq:5-01} that all theta terms can be expressed as four-divergences which do not affect the classical equations of motion.  Their physical effects are non-perturbative and come from instantons which depend on topologically nontrivial vacuum configurations of gauge fields.

The first term of~\eqref{eq:5-01} is the well-known QCD theta term.  Its coefficient $\theta_3$ also appears in the expression~\eqref{eq:4-03} for $\bar \theta_3$, which can be identified as the effective QCD theta angle $\bar \theta_\text{QCD}$ and leads to physical observables such as the neutron EDM~\cite{Baluni:1978rf, Crewther:1979pi, Pospelov:1999ha, Pospelov:1999mv, Liu:2024kqy}.  One can apply a chiral rotation with~\eqref{eq:4-10} to remove $\theta_2$, so the $SU(2)$ theta angle is not a physical parameter of the SM, unless beyond the SM operators are considered to prevent the removal of $\theta_2$~\cite{Anselm:1992yz, Anselm:1993uj, FileviezPerez:2014xju, Shifman:2017lkj}.  Terms proportional to $\theta_1$ in~\eqref{eq:5-01} contain field strengths of $A_\mu$ and $Z_\mu$.  If these fields are configured on a compact and non-simply connected spacetime manifold~\cite{Witten:1995gf, Verlinde:1995mz, Olive:2000yy}, the energy scale of their instanton effects is set by the size of the compact manifold, and exponentially suppressed in $1 / e^2$.  Since $Z_\mu$ gets a nonzero mass after EWSB, its instanton effect has an additional power-suppression in the $Z_\mu$ mass, which is proportional to the vacuum expectation value of the Higgs field~\cite{tHooft:1976snw}.  In addition, it is easier to configure in a laboratory a topologically nontrivial background with $A_\mu$ than with $Z_\mu$.  Therefore the leading contribution to the physical effect of $\theta_1$ comes from the second term of~\eqref{eq:5-01}, which only contains the field strength of the massless electromagnetic field $A_\mu$.  Such a laboratory construction with nontrivial topology can be effectively realized in an interferometer setup analogous to the Aharonov-Bohm experiment but with a nonzero magnetic helicity~\cite{Hsu:2010jm, Hsu:2011sx}, or in a parallel conductor plate system with a uniform magnetic field background~\cite{Cao:2017ocv, Zhitnitsky:2023jvp}.  It may also lead to cosmological consequences by considering non-simply connected features beyond the visible universe~\cite{Cao:2013na}.

The second term of~\eqref{eq:5-01} has the same form as a QED theta term.  Its coefficient $\frac{1}{2} \theta_2 + \theta_1$ also appears in the expression~\eqref{eq:4-04} for the invariant combination $\bar \theta_{2 1}$.  While the coefficients of the third, fourth and fifth terms of~\eqref{eq:5-01} can not be removed simultaneously by a chiral rotation, their possible physical effects, compared to those of the second term, are power-suppressed in the masses of $Z_\mu$ and $W^\pm_\mu$.  In addition, it is easier to conduct an experiment on the QED effects of $\bar \theta_{2 1}$ with the massless electromagnetic field $A_\mu$ compared to those with the massive gauge fields.  Thus $\bar \theta_{2 1}$ can be identified as the effective QED theta angle $\bar \theta_\text{QED}$ which is also invariant under arbitrary chiral rotations of fermions.  Unlike $\bar \theta_\text{QCD}$ whose physical effects are proportional to the order-one factor $\exp (- 8 \pi^2 / g_3^2)$ and present in Minkowski spacetime, $\bar \theta_\text{QED}$ has physical effects suppressed by $\exp (- 8 \pi^2 / e^2)$ and present only in spacetime with non-simply connected features, either beyond the visible universe or in an effective background from a laboratory setup.  We conclude that $\bar \theta_\text{QED}$ is the physical remnant of the SM $SU(2)$ and $U(1)$ theta angles after EWSB, and should be viewed as an independent parameter of the SM.

Although~\eqref{eq:4-04} and~\eqref{eq:5-01} contain the same combination $\frac{1}{2} \theta_2 + \theta_1$, they are from different origins.  The coefficients in~\eqref{eq:4-04} come from canceling the effects of chiral rotations on theta angles and phases of fermion mass matrices, which depend on the gauge group representations of fermions.  On the other hand, the coefficients $\frac{1}{2}$ and $1$ of $\frac{1}{2} \theta_2 + \theta_1$ in~\eqref{eq:5-01} are the second-order indices $I^{(2)}_F$ and $I^{(1)}_F$ for fundamental representations of $SU(2)$ and $U(1)$, which come from traces over $F^{(2) a}_{\mu \nu} F^{(2) a}_{\kappa \lambda}$ and $F^{(1) a}_{\mu \nu} F^{(1) a}_{\kappa \lambda}$ in \eqref{eq:3-01}.  This coincidence could provide constraints on the gauge group representations of SM fermions, which may be worth further exploration.

\section*{Acknowledgement}

The authors thank Andrea Addazi, Yan He, Huiji Jin, Chengyang Lee, Jinmian Li, Takaaki Nomura and Zhiguang Xiao for helpful discussions.  This work is supported by the National Natural Science Foundation of China (NSFC) under the grant numbers 12205208, 12275185 and 12335002.

\end{document}